\documentclass[article]{elsarticle}
\usepackage[T1]{fontenc}
\usepackage{lineno,hyperref}
\modulolinenumbers[5]

\journal{JQSRT doi:10.1016/j.jqsrt.2018.02.020}

\begin{document}

\begin{frontmatter}
	
	\title{Experimental and theoretical   study of the B(2)$^{2}\Sigma^{+}\rightarrow$ X(1)$^{2}\Sigma^{+}$   system in the KSr molecule}
		
\author{Jacek Szczepkowski \corref{mycorrespondingauthor}}
\cortext[mycorrespondingauthor]{Corresponding authors}
\ead{jszczep@ifpan.edu.pl}
\author{Anna Grochola  \corref{c2}}

\address{Institute of Physics, Polish Academy of Sciences, Al. Lotnik\'ow 32/46,
	02-668 Warszawa, Poland}
\author{Pawe\l{} Kowalczyk \corref{c3}}
\address{ Institute of Experimental Physics, Faculty of Physics, University of Warsaw, ul.~Pasteura~5, 02-093 Warszawa, Poland}

\author{Olivier Dulieu* \corref{c4}}

\ead{olivier.dulieu@u-psud.fr}
\address{Laboratoire Aim\'{e} Cotton, CNRS, Universit\'{e} Paris-Sud, ENS Paris-Saclay, Universit\'e Paris-Saclay, B\^at. 505, rue Aim\'e Cotton, F-91405 Orsay Cedex, France}
\author{Romain Gu\'erout \corref{c41}}
\address{Laboratoire Kastler Brossel, CNRS, ENS, UPMC, Case 74,F-75252, Paris, France}

\author{Piotr S. \.Zuchowski* \corref{c5}}
\ead{pzuch@fizyka.umk.pl}
\address{Institute of Physics, Faculty of Physics, Astronomy and Informatics, Nicolas Copernicus University, ul. Grudzi\k{a}dzka 5/7, 87-100 Toru\'{n}, Poland}

\author{W\l{}odzimierz Jastrzebski \corref{c1}}
\address{Institute of Physics, Polish Academy of Sciences, Al. Lotnik\'ow 32/46,
	02-668 Warszawa, Poland}
	

\begin{abstract}
Spectral bands for the B(2)$^{2}\Sigma^{+}\rightarrow$ X(1)$^{2}\Sigma^{+}$ electronic transition in the doubly-polar open-shell KSr molecule are recorded at moderate resolution using the thermoluminescence technique. The spectra are simulated using three kinds of advanced electronic structure calculations, allowing for an assessment of their accuracy on one hand, and for the derivation of fundamental spectroscopic constants of the X(1)$^{2}\Sigma^{+}$ KSr ground state and the excited electronic state B(2)$^{2}\Sigma^{+}$, on the other hand. These results should facilitate further studies aiming at creating ultracold bosonic or fermionic KSr molecules.

	\end{abstract}
	

\end{frontmatter}


\section{Introduction}

In recent years mixed diatomic molecules composed with one alkali-metal atom (Li, Na, K, Rb, Cs) and one alkaline-earth atom (Be, Mg, Ca, Sr, Ba)  became particularly interesting for the community of physicists, due to the possibility of performing new experiments in the field of ultracold molecules \cite{Perez-Rios2010,Chen2014,Quemener2016,Barbe2017}. Such open-shell diatomics are called "doubly polar molecules" since they have both electric and magnetic permanent dipole moments. At ultralow temperatures they provide unique opportunities for quantum control and new possibilities for quantum simulations \cite{Micheli2006,Baranov2012,Lahaye2009}. Such challenging experiments require guidance based on experimentally known features of the chosen species. However, because of experimental difficulties, only a limited number of studies have been published so far. Low-resolution molecular spectroscopy - complemented by advanced electronic structure calculations - has been achieved on RbSr \cite{Lackner2014,Krois2014,Pototschnig2014}, RbCa \cite{Pototschnig2015}, and LiCa \cite{Krois2013} molecules embedded inside helium nanodroplets as cold as a few kelvins, thus drastically reducing the number of populated levels of the electronic ground state X(1)$^{2}\Sigma^{+}$ (typically the lowest vibrational state, and a few rotational states). The explored spectral range concerned the excited molecular states above the two lowest ones, namely (2)$^{2}\Sigma^{+}$ and (2)$^{2}\Pi$. 

In contrast, high-resolution studies based on Fourier transform spectroscopy and addressing the latter excited electronic states have been recently reported for LiSr \cite{Schwanke2017} and KCa \cite{Gerschmann2017}, and on LiBa in a series of older papers \cite{Verges1994,Stringat1994,DIncan1994}. It is worthwhile also to note the resonant two-photon ionization spectroscopy of LiCa reported in Ref.~\cite{russon1998}.

Despite the apparent simplicity of alkali--alkaline-earth diatomic molecules, the accurate computation of their electronic structure is cumbersome when heavy atoms are involved, as strong correlations are expected between the valence electrons and those of the closed-shell ionic cores. Numerous exploratory calculations have been performed over the years with almost no possibility to evaluate them against experiment, up to very recently. We can quote papers on barium-alkali-metal-atom compounds \cite{Allouche1994,boutassetta1994,boutassetta1995,Gou2015}, lithium-alkaline-earth-atom compounds \cite{kotochigova2011}, strontium-alkali-metal-atom compounds \cite{Guerout2010,Zuchowski2014}, and other combinations \cite{Gopakumar2013} up to the recent systematic investigation for sixteen alkali--alkaline-earth compounds of Ref.~\cite{Pototschnig2016}
 
Among the alkali-metal--alkaline-earth species, KSr definitely belongs to the most interesting ones. The potassium atom is the only alkali-metal atom with three stable or long-lived isotopes, which together with four isotopes of Sr give 12 possible mass combinations exhibiting both bosonic (e.g. $^{39}$K+$^{87}$Sr) and fermionic (e.g. $^{40}$K+$^{87}$Sr) statistics. Such a variety of reduced masses might be used to tune the scattering properties in KSr by proper selection of the isotopologue. 

It is well known that the accuracy of quantum chemistry calculations does not generally meet  the requirements of ultracold physics. For instance the scattering length for a pair of colliding atoms cannot be predicted without the help of high-resolution experimental data. It is also challenging to approach the spectroscopic accuracy for the ground state and the excited electronic states. In most cases it is essential to combine experimental data and theoretical potential energy curves to derive an ''experimental'' potential energy curve. On the other hand, experimental techniques like Laser Induced Fluorescence (LIF) or Polarisation Labelling Spectroscopy (PLS), which yield spectroscopic data with rotational resolution, must be initiated with accurate theoretical data, or with data delivered by less demanding experiments, as it was shown in case of KCa molecule \cite{Gerschmann2017}. The thermoluminescence technique used in the present work belongs to this latter category of experimental methods, thus representing an important step to refine potential energy curves.       

The present paper originates from a close cooperation between theoretical and experimental groups. We first present in Section \ref{sec:theory} the electronic structure calculations performed with two different methods, in order to assess their typical accuracy through the comparison of their results. In Section \ref{sec:experiment} we describe our experimental approach based on the thermoluminescence technique, which allowed us to extract the first experimental results for the X(1)$^{2}\Sigma^{+}$ and B(2)$^{2}\Sigma^{+}$ electronic states of the KSr molecule. Through the simulations of the recorded spectra (Section \ref{sec:simulations}), we demonstrate in Section \ref{sec:analysis} that the level of accuracy of the present calculations allowed a better understanding and a proper interpretation of the spectroscopic data. In the future the relatively precise experimental molecular constants should be used to correct the assumptions of the individual theoretical models and to improve the accuracy of the predicted molecular parameters.  

\section{Electronic structure calculations}
\label{sec:theory}

As quoted above, the valence electrons (one for K, two for Sr) and the remaining core electrons of K$^+$ and Sr$^{2+}$ in the KSr system strongly interact, making the computation of their electronic structure tedious. Therefore the usage of different methods to perform such calculations is mandatory for a documented assessment of their validity. We proceeded here like in our previous collaborative investigations, employing two different methods. Below we recall their main steps, for reader's convenience, while all details and references can be found in Ref.~\cite{Zuchowski2014}.

The first approach, for simplicity hereafter referred to as the FCI+ECP+CPP method, relies on the modelling of the K$^+$ and Sr$^{2+}$ closed-shell cores by an effective core potential (ECP) completed by a core-polarization potential (CPP) \cite{muller1984,muller1984a,foucrault1992}, which parameters are reported in Ref.~\cite{Guerout2010}. The Gaussian orbital basis sets spanning the configuration space are also reported in Ref.~\cite{Guerout2010}. Thus the problem is reduced to an effective three-electron system, allowing the resolution of the electronic Schr\"odinger equation using a full configuration interaction (FCI). This yields the potential energy curves (PECs) of the electronic ground state X$^2\Sigma^+$, as well as many excited electronic states of $^{2,4}\Sigma^+$, $^{2,4}\Pi$, and $^{2,4}\Delta$ symmetries, and subsequently the permanent and transition dipole moments (PDM and TDM) involving these states as a function of the internuclear distance $R$.

The second method employs 19 active electrons whereas the inner electronic shells are treated as effective core potential ECP referred to as ECP28MDF for Sr and ECP10MDF for K, so that the $4s$, $4p$, and $5s$ electrons of Sr, and the $3s$, $3p$ and $4s$ electrons of K are fully correlated. The ECPs used in this calculations were obtained by Lim et al. \cite{Lim2005,Lim2006} and the basis set tailored for this ECP was uncontracted and augmented by adding one $h$ function to the basis set for each atom (with exponents  3.84 and 0.56 for K and Sr, respectively). The basis set was subsequently augmented by a set of diffusion functions. The midbond basis set was added with exponents 0.9,0.3,0.1 for $spd$-type functions and 0.6,0.2 for $fg$ functions, and placed on ghost atom in geometric centre of the molecule. The PECs are obtained using counterpoise-corrected supermolecular coupled-cluster method based on spin-restricted reference \cite{knowles1993} with single and double excitations and non-iterative correction for triply-excited configurations [RCCSD(T)]. For the ground-state closed-shell system, this method is considered as the golden standard of the quantum chemistry calculations of interaction potential. However, for the ground state open-shell systems RCCSD(T) were studied far less in terms of uncertainty of calculations of PECs compared to closed-shell systems. For the first excited states, the calculations of potential energy curves which employ such approach are  explored very little (note for example the work of Klos et al.~\cite{Kolos2008}.  In this respect it is very appealing to confront RCCSD(T) method applied to the ground state and the first excited $^2\Sigma^{+}$ state, and compare it with alternative methods. Such a comparison will be profitable for  future studies of similar systems relevant in ultracold molecular physics. While the former state is easy to converge and maintains the single-reference character for broad range of distances, the supermolecular RCCSD(T) calculation for the second $^2\Sigma^{+}$ state was a bit more demanding. To this end, the orbital rotation of $4s$ and $4p$ potassium orbitals were applied on top of the ground state reference state, and subsequently converged.
The coupled-cluster equations could be easily converged down to about 2.5$\AA$. It is worth noting that the value  of T1 diagnostics of Lee \cite{Lee2003} used commonly for  assessment of multireference character of CCSD calculation is at most 0.03 in the energy well, slightly more than the value indicating single-reference character. The {\sc molpro 2012 } package was used for this part of {\em ab initio} calculations \cite{molpro2012}.

In both methods, we model the long-range part of the potential curves using the standard $-C_6 / R^{6}- C_8/ R^{8}$ expansion smoothly matched to the curves obtained above (at 30~a.u. for the FCI+ECP+CPP method, and at 20~a.u. for the RCCSD(T) method) with switching function taken from Ref. \cite{Janssen2009}. The van der Waals coefficients were taken from the paper of Jiang et al. \cite{Jiang2013}. 
 
\begin{figure}
	\includegraphics[width=0.9\linewidth]{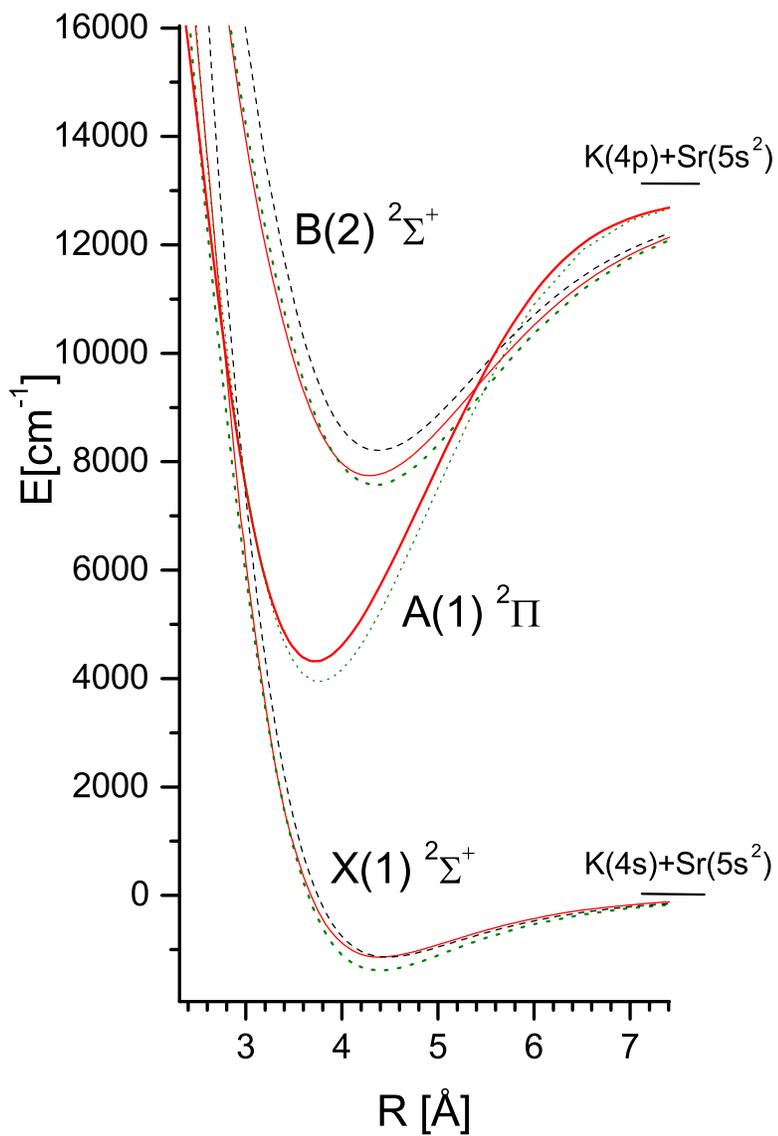}
	\caption{Comparison of three sets of theoretical potential energy curves for the 
		ground and two lowest exited  states of KSr. Results are displayed for the FCI+ECP+CPP approach (red solid lines), for the RCCSD(T) approach (black dashed lines), and for the MRCI potentials of the Graz group \cite{Pototschnig} (green dotted lines). The original energy scales used in the calculations, locating zero of energy at the dissociation limit of the ground state, are preserved.}
	\label{fig:KSrtheory}
\end{figure}

\begin{figure}
	\includegraphics[width=0.9\linewidth]{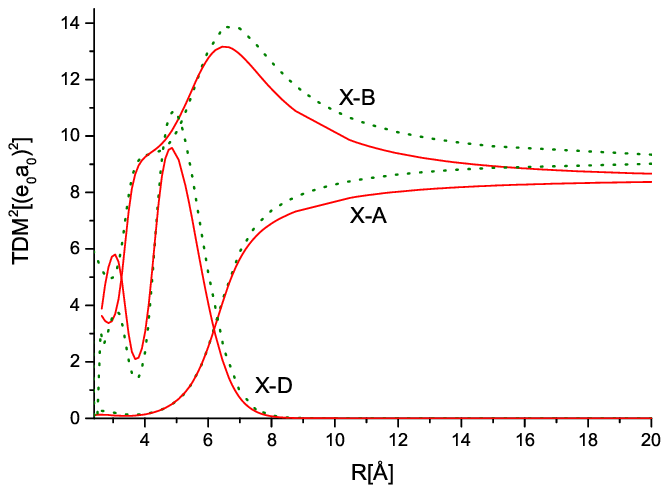}
	\caption{The transition dipole moments for the $X(1)^{2}\Sigma^+$--$A(1)^{2}\Pi$, $X(1)^{2}\Sigma^+$--$B(2)^{2}\Sigma^+$ and  $X(1)^{2}\Sigma^+$--$D(3)^{2}\Sigma^+$ transitions in the KSr molecule. Results are displayed for the FCI+ECP+CPP approach (red solid lines), and for the MRCI results (green doted lines) of the Graz group \cite{Pototschnig}.}
	\label{fig:KSrtheoryTDM}
\end{figure}

The results for the PECs of the $X(1)^2\Sigma^+$ ground state and the lowest $A^{2}\Pi$ and $B^{2}\Sigma^+$ states relevant for the present study are displayed in Fig. \ref{fig:KSrtheory} and the related transition dipole moment (TDM) in Fig. \ref{fig:KSrtheoryTDM}, together with the results reported previously in Ref. \cite{Pototschnig} (by the group henceforth referred to as the Graz group) based on the multireference configuration interaction (MRCI) approach. 

The agreement between the RCCSD(T) approach and the FCI+ECP+CPP method for the ground state is remarkable, and confirms the trend already reported in Ref.~\cite{Zuchowski2014} for RbSr. In contrast, the MRCI approach of Ref. \cite{Pototschnig} predicts a significantly deeper potential well with a rather strong harmonic constant (see Tab. \ref{tab:KSrconstants}). As the RCCSD(T) approach is currently considered as the most reliable method for such ground state calculations, an as its results are confirmed by a completely different semi-empirical approach, we recommend these results for further use. The situation is less clear for the lowest excited electronic states: all methods provide consistent results, but again the well depth reported in Ref. \cite{Pototschnig} is significantly deeper than in the two methods. 

\begin{figure}
	\includegraphics[width=0.9\linewidth]{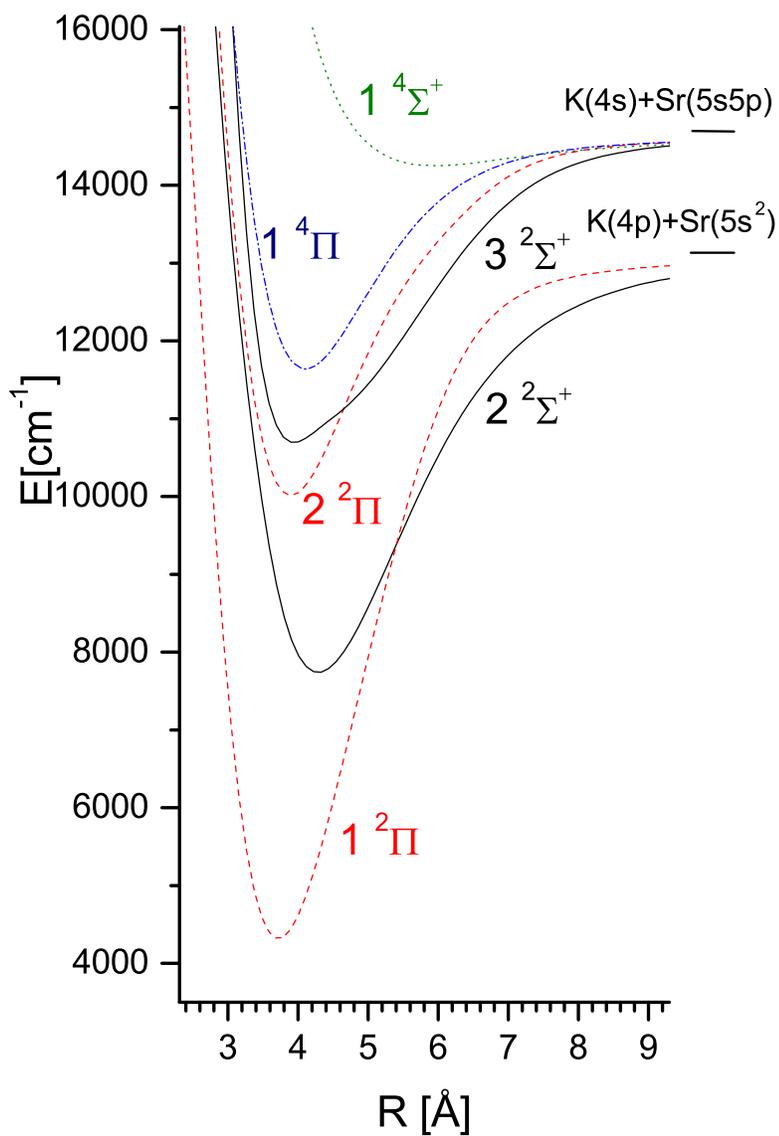}
	\caption{The potential energy curves calculated with FCI+ECP+CPP method for all states correlating to the K(4p$\;^2P$)+Sr(5s$^2\;^1S$) and K(4s$\;^2S$)+Sr(5s5p$\;^3P$) asymptotes.	 The solid black lines denote $^{2}\Sigma^+$ states, the dashed red lines denote $^{2}\Pi$, the dash-dot blue line refers to $^{4}\Pi$ and the dotted green line denotes 
		$^{4}\Sigma^+$ state. The zero energy  is the same as in Fig.\ref{fig:KSrtheory}  }
	\label{fig:KSrtheoryFCI}
\end{figure}

The TDM is well known as a quantity very sensitive to the details of the electronic wave functions. The agreement between the FCI+ECP+CPP method and the one of the Graz group is remarkable (see Fig.\ref{fig:KSrtheoryTDM}) especially in the 4\AA-6\AA~range, where the oscillation for the B state originates from a broad avoided crossing-as large as about $1500\,cm^{-1}$ between the $B(2)^{2}\Sigma^+$ and $D(3)^{2}\Sigma^+$ excited states (see Fig.\ref{fig:KSrtheoryFCI}). Such an agreement clearly expresses that both methods, while very different, describe the electronic wave functions in the same way, thus representing an indication of their validity. However it is striking that the asymptotic values of the B-X and A-X TDM reach the atomic  K($4s\;^2S$)-K($4p\;^2P$) TDM in the FCI+ECP+CPP approach, while it is overestimated by  8~\% in the results of the Graz group.

\begin{table}
	
	\caption{
		Spectroscopic constants of the KSr system obtained from theoretical calculations. The vibrational constants 
		correspond to $^{39}$K$^{88}$Sr  isotopic combination. See text for the details.}
	\begin{tabular}{cccccc}\hline\hline
		state  & $T_e\:[cm^{-1}]$   & $D_e\:[cm^{-1}]$        &      $\omega_e\:[cm^{-1}]$ & $B_e\:[cm^{-1}]$ & $R_e$ [\AA{}] \\ \hline
		\multicolumn{6 }{c}{RCCSD(T) (present paper) }  \\ \hline
		$X(1)^2\Sigma^+$ &  0 &     1138.56     &     52.27&  0.0312&   4.47\\
		$B(2)^2\Sigma^+$ &   9346.15 &     4845.46     &     74.27&0.0327 &  4.37 \\ \hline
		\multicolumn{6 }{c}{FCI+ECP+CPP  (present paper)   }  \\ \hline
		$X(1)^2\Sigma^+$ & 0  &     1164.67     &     52.82& 0.0322&    4.40\\
		$B(2)^2\Sigma^+$ & 8905.98 &     5282.15     &     77.31&0.0339 &    4.29\\ \hline
		\multicolumn{6 }{c}{ MRCI (Pototschnig et al.\cite{Pototschnig})  }  \\ \hline
		$X(1)^2\Sigma^+$ &  0  &     1383.41    &     56.49& 0.0324&   4.39\\
		$B(2)^2\Sigma^+$ &  8954.27 &     5507.32     &     76.18&  0.0329&   4.35 \\ \hline
	\end{tabular}
	\label{tab:KSrconstants}
\end{table}

The data for the PECs and TDMs computed in the present work are reported in the Supplementary material \cite{supplC}.  Note that according to our calculations with the FCI+ECP+CPP method,  the PECs correlated to the next dissociation limit, K(4s$\;^2S$)+Sr(5s5p$\;^3P$), have all their minima well above the one of the PECs correlated to K(4p$\;^2P$)+Sr(5s$^2\;^1S$) studied here (see Fig.\ref{fig:KSrtheoryFCI}). Beside the $D(3)^2\Sigma^+$ PEC  quoted above, the other PECs $(1)^{4}\Sigma^+$, $(2)^{2}\Pi$ and $(1)^{4}\Pi$ do not cross the one of the B and A state, and thus are not expected to play a role in the present results. Moreover, the minimum of the $(1)^{4}\Pi$ PEC is predicted around $10500\,cm^{-1}$, thus largely separated from the B$(2)^{2}\Sigma^+$ PEC, and should not  significantly contribute to the spin-orbit energy in the B state (see Fig.\ref{fig:KSrtheoryFCI}).

\section{The thermoluminescence experiment}
\label{sec:experiment}

At temperatures high enough a thermal population of electronically excited molecular states is possible. Since working temperatures for potassium and strontium are significantly different, a special dual-temperature stainless steel heat-pipe oven was constructed to produce a vapor of KSr molecules. A detailed description of this device can be found in Ref.~\cite{Bednarska1996}. In general, the heat-pipe was spatially divided into three sections. The central part of the pipe was filled with 10~g of strontium and heated to the temperature $T_{Sr}$$\;$=$\;$1100$\;$K, while the outer parts were filled with 10~g of potassium and maintained at the temperature $T_{K}$$\;$=$\;$820$\;$K. Only potassium parts of the oven were covered inside with a stainless steel mesh to provide a proper metal circulation in the heat-pipe. Helium at pressure of $30$~Torr was used as a buffer gas to assure stability of the heat-pipe operation and to prevent metal deposition on the oven windows. The atomic vapors were mixing and reacting in the central part of the heat-pipe, producing KSr molecules. Thermally excited fluorescence spectra were recorded using Bruker Vertex V80 Fourier Transform Spectrometer with a spectral resolution 0.16$\;$cm$^{-1}$, limited by an aperture size. 

\begin{figure}
	\includegraphics[width=0.9\linewidth]{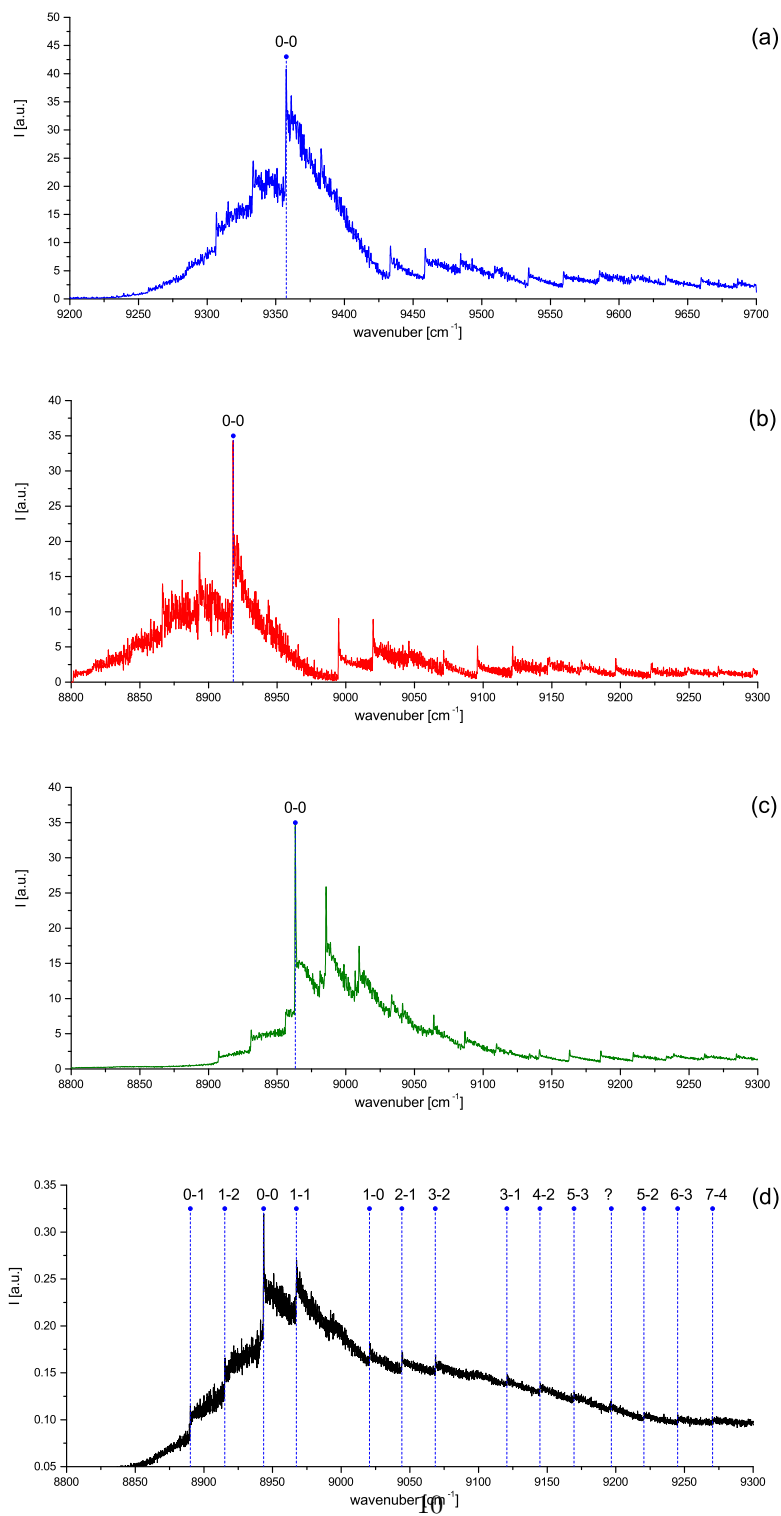}
	\caption{Comparison of the B(2)$^{2}\Sigma^{+}$ $\leftarrow$ X(1)$^{2}\Sigma^{+}$ spectra in the KSr molecule simulated using theoretical potential curves from the present paper (obtained by RCCSD(T) method (a) and by FCI+ECP+CPP approach (b)), and from Ref.~\cite{Pototschnig} (c) with the experimental spectrum from this work (d). The assignment of band heads is based on theoretical calculations (vertical dashed line).}
	\label{fig:expsim}
\end{figure}

At the applied temperatures, the KSr second excited state, namely the B(2)$^{2}\Sigma^{+}$ state, (the first excited state being the A(1)$^{2}\Pi$ state)  was populated thermally, and fluorescence to the ground state was recorded. The population of the B(2)$^{2}\Sigma^{+}$ state rovibrational levels was determined by the temperature of the KSr molecules, particularly of the central part of the heat-pipe, according to the Boltzmann distribution. The band heads corresponding to transitions between different vibrational levels of the B(2)$^{2}\Sigma^{+}$ and X(1)$^{2}\Sigma^{+}$ electronic states were clearly seen (see Fig.\ref{fig:expsim}d). At the resolution of the present experiment, each unresolved band head consists of a group of about 30 rotational lines.

\section{Simulations of the recorded spectra}
\label{sec:simulations}

Simulations of molecular spectra were performed using theoretical potentials in order to compare them with the experimental results. Such comparison was necessary to enable the correct assignment of the band heads and further analysis of the experimental data. Calculations were based on three sets of theoretical potential energy curves, generated independently with FCI+ECP+CPP and RCCSD(T) methods (see Section \ref{sec:theory}) as well as originating from the Graz group \cite{Pototschnig}.

In the first step of the simulation energies of rovibrational levels of the B(2)$^{2}\Sigma^{+}$ and X(1)$^{2}\Sigma^{+}$ states were calculated by solving the radial Schr\"{o}dinger equation with theoretical potentials. All bound levels in the X(1)$^{2}\Sigma^{+}$ state were taken into account, spanning the range $v''$=0--55, $N''$=0--217 (the upper value corresponding to the highest bound rotational level for $v''$=0), where $N$ denotes the quantum number of the total angular momentum apart from spin, being the natural rotational quantum number for $^2\Sigma^{+}$ levels. The highest vibrational level to be considered in the B(2)$^{2}\Sigma^{+}$ state was determined by calculated Franck-Condon factors (FCF). We included in the simulations all B(2)$^{2}\Sigma^{+}$ vibrational levels with non-zero FCF values for transitions to the ground state levels (i.e. $v'$=0--80). Note that we checked that the contribution of the A(1)$^{2}\Pi$ state to the observed spectra is found typically 2000 times smaller than the one of the B state, due to the combined effect of smaller TDM and unfavorable FCFs.

Several assumptions were made to simulate the spectra. The fine structure splitting of molecular levels (i.e. splitting between F$_{1}$ and F$_{2}$ levels due to the spin-rotation coupling) was neglected, since the distance between fine structure components for low rotational quantum numbers $N'$, for which the band heads are formed, is smaller than the spectral resolution. The intensities of the transitions B($v''$,$\;$$N''$)$\;$$\rightarrow$$\;$X($v''$,$\;$$N''$) were calculated assuming thermal equilibrium, according to the formula
\begin{equation}
I_{trans}(v',N'\rightarrow v'',N'') \sim HL(N'\rightarrow N'') \cdot P(v',N')  \cdot \left| \left\langle \psi'_{v',N'}|\mu|\psi''_{v'',N''} \right\rangle \right|^{2}  ,
\end{equation} 
where $\left| \left\langle \psi'_{v',N'}|\mu|\psi''_{v'',N''} \right\rangle \right|^{2}$ corresponds to FCF including theoretical transition dipole moment $\mu$ (taken either from the present calculation or from the Graz group \cite{Pototschnig}, see Fig.\ref{fig:KSrtheoryTDM}), $HL$($N'$$\;$$\rightarrow$$\;$$N''$) denotes normalized H\"{o}nl-London factors for P and R branches given by the formula \cite{Ochkin2009}
\begin{equation}
HL_{P}(N')=\frac{1}{2(2N'+1)}\left(N'-\frac{1}{4N'}\right) ,
\end{equation}
\begin{equation}
HL_{R}(N')=\frac{1}{2(2N'+1)}\frac{(N'+1/2)(N'+3/2)}{(N'+1)},
\end{equation}
and $P$($v', N'$) describes population of energy levels in the B(2)$^{2}\Sigma^{+}$ state at thermal equilibrium
\begin{equation}
P(v',N') = 2(2N'+1)\exp \left(-\frac{B_{e}^{B}N'(N'+1)}{kT}\right)\exp \left(-\frac{T_{e}^{B}+\omega_{e}^{B}(v'+1/2)}{kT}\right).
\end{equation}
The values of spectroscopic parameters $T_{e}$, $\omega_{e}$ and $B_{e}$ of the excited state were taken from the theoretical predictions. The line profiles were assumed to have a Gaussian shape with FWHM$\;$=$\;$0.16$\;$cm$^{-1}$, that results from the Fourier Transform Spectrometer working parameters. The simulated spectra were used in the process of identification of the experimental results.

\section{Data Analysis}
\label{sec:analysis}

A specific method of experimental spectra analysis was used. At the beginning three theoretical spectra of the B$\rightarrow$X transition were simulated according to the procedure described in the previous section, relying on three sets of potential curves, either from Ref.~\cite{Pototschnig} or from the present calculations. These simulated spectra were compared with the experimental spectrum, as shown in Fig.\ref{fig:expsim}. It can be noticed that the general shape of the spectra is similar and the main difference comes from different theoretical values of term energy of the $B$ state ($T_e^B$) in each calculation(see Tab.\ref{tab:KSrconstants}), resulting in different positions of the spectra on the energy (wavenumber) scale. The relative positions of the individual band heads and their intensities are also somewhat different for the three theoretical calculations of interest. Nevertheless, by careful comparison, we were able to assign several band heads in the experimental spectrum, as shown in Fig.\ref{fig:expsim}(d). The wavenumbers of the assigned band heads can be arranged neatly into a Deslandres table (Table~\ref{tab:deslandres_exp_KSr}), which supports our assignment. 

\begin{table} 
	\caption{Deslandres table constructed for the observed band heads in the experimental spectrum of KSr.}
	\label{tab:deslandres_exp_KSr}
	\begin{tabular}{ccccccccc}
		\hline
		\hline
			      & $v''=0$    &         & $1$ &     & $2$ &   & $3$ &     $4$ \\ \hline
		$v'=0$ & $8943.67$ & $53.54$ & $8890.13$ &   &   &   &   &    \\ 
		& $77.21$   &         & $77.38$ &   &   &   &   &    \\ 
		$1$ & $9020.88$ & $53.37$ & $8967.51$ & $52.31$ & $8915.2$ & & &  \\ 
		&           &         & $76.08$ &  &  &  &  &   \\
		$2$ &           &         & $9043.59$ &   &   &   &   &    \\ 
		&           &         & $77.33$ &   &   &   &   &    \\ 
		$3$ &           &         & $9120.92$ & $52.18$ & $9068.74$ & & &  \\
		&           &         & $75.51$ &           & $76.27$ & &  & \\ 
		$4$ &           &         & $9196.43$ & $51.42$ & $9145.01$ &  & & \\ 
		&           &         &           &         & $75.08$ & &  & \\ 
		$5$ &  &  &  &  & $9220.81$ & $51.18$ & $9169.63$ &  \\ 
		&     &     &     &     &    &    & $75.48$ &  \\ 
		$6$ &     &     &     &     &    &    & $9245.11$ &  \\ 
		$7$ &   &   &   &   &   &   &      & $9270.29$ \\ \hline
	\end{tabular}
\end{table}

\begin{figure}[h]
	\includegraphics[width=0.9\linewidth]{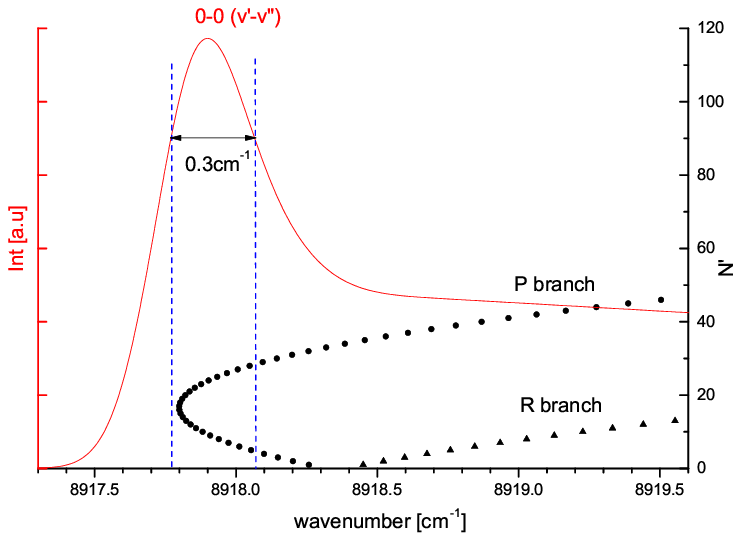}
	\caption{An example of band head formation in the 0--0 band of the B--X system of KSr. Vertical dashed lines indicate the range of rotational levels involved in the band head formation and assumed to correspond to the same wavenumber in further analysis.}
	\label{fig:bandheadformation}
\end{figure}

At the next stage of the analysis we ascertained which rotational transitions were responsible for formation of each band head, taking into account the experimental width of the band heads, typically 0.3~cm$^{-1}$ (see Fig.\ref{fig:bandheadformation}).
\noindent All the observed vibrational bands were red shaded which means that the band heads were formed in the P branches, the R branches providing merely a background in the recorded spectra. (This observation implies that the equilibrium distance $R'_e$ of the $B$ state is smaller than $R''_e$ of the ground state). A range of the $B$ state rotational levels $N'$ involved in formation of each band head was identified on the basis of theoretical calculations and only transitions starting from these levels were taken into account in further steps of the analysis. Note that a different range of $N'$ values was attributed not only to each band head but also to each theoretical model. In this way a group of P lines responsible for band head formation was specified for each band and we assigned to these lines the same wavenumber, namely that of the corresponding band head. As a result we obtained a set of about 350 P lines, split into 13 groups, with the same wavenumber within each group. The wavenumbers of these lines were fitted to differences of term values of the upper and lower levels,
\begin{equation}
\nu=T_B(v',N')-T_X(v'',N'') \mbox{  ,}
\end{equation}
\noindent which were expressed by a standard Dunham expansion 
\begin{equation}
T_{B}=T_e^{B}+Y_{10}^{B}(v'+0.5)+Y_{01}^{B}(N'(N'+1))+Y_{20}^{B}(v'+0.5)^{2}+Y_{11}^{B}(v'+0.5)(N'(N'+1)),
\end{equation}
\begin{equation}
T_{X}=Y_{10}^{X}(v''+0.5)+Y_{01}^{X}(N''(N''+1))+Y_{20}^{X}(v''+0.5)^{2}+Y_{11}^{X}(v''+0.5)(N''(N''+1)),
\end{equation}
\noindent with the Dunham coefficients $Y_{ij}$ having their usual meaning. The value of the rotational constant of the ground state $Y_{01}^{X} \equiv B_e^{X}$ had to be taken from theoretical predictions because of the unresolved rotational structure of the experimental thermoluminescence spectra. Three sets of Dunham coefficients were generated for both the $B(2)^{2}\Sigma^{+}$ and the $X(1)^{2}\Sigma^{+}$ states assuming three different values of rotational constant $Y_{01}^{X}$ resulting either from the present calculations or these from Ref. \cite{Pototschnig} (Table~\ref{tab:KSrDunham}). Comparison of the three sets of coefficients for each state shows that their values (except of the enforced values of rotational constants $Y_{01}$) agree within their uncertainties despite significant differences between theoretical models. Therefore one can expect that the obtained Dunham coefficients may be used to predict fairly accurately values of energies of vibrational levels in the $X(1)^{2}\Sigma^{+}$ and $B(2)^{2}\Sigma^{+}$ states also beyond the range of this experiment.
Finally we note that the $Y_{10}^{X} \equiv \omega_e^{X}$ values in Table~\ref{tab:KSrDunham} lie in between the one of the present calculations, and the one from Ref. \cite{Pototschnig}. In contrast, the obtained value for $Y_{10}^{B} \equiv \omega_e^{B}$ is larger than the theoretical values from RCCSD(T) and the Graz results, while it is in remarkable agreement with the value from the FCI+ECP+CPP method. The $T_e$ value derived from the Dunham analysis is also in good agreement with the results of the FCI+ECP+CPP calculation and that of Ref.~\cite{Pototschnig}, and smaller than the one obtained by the RCCSD(T) approach. It is probable that this large deviation of $T_e$ in RCCSD(T) results from the strong multireference character of the B state - which exhibits a marked  avoided crossing with the next electronic state, a quoted above - that is not accounted for in the coupled-cluster method. 

\begin{table}
	\caption{The experimental Dunham coefficients for the B(2)$^{2}\Sigma^{+}$ and X(1)$^{2}\Sigma^{+}$ states of the KSr molecule. The three different sets of coefficients for each state rely on different values of the ground-state equilibrium distance $R_e$ taken from three theoretical calculations. All values are in cm$^{-1}$.}
	\label{tab:KSrDunham}
	\begin{tabular}{cccc}	
	\hline
	\hline
$Y{i,j}$	& MRCI \cite{Pototschnig}& FCI+ECP+CPP& RCCSD(T)\\ \hline
	&\multicolumn{3}{c}{$X (1)^{2}\Sigma^{+}$}\\
	\multicolumn{1}{c}{$Y_{10}$} & $54.83 \pm 1$ & $54.58 \pm 1$& $54.59 \pm 1$\\
	\multicolumn{1}{c}{$Y_{20}$ }& $-0.59 \pm 0.19$& $-0.56 \pm 0.18$& $-0.56 \pm 0.18$ \\
	\multicolumn{1}{c}{$Y_{01}\times 10^{2}$} & $3.23557^{a}$&$3.22434^{a}$&$3.12098^{a}$\\
	\multicolumn{1}{c}{$Y_{11}\times 10^{4}$} & $-2.2 \pm 4$& $-2.7 \pm 5$ & $-2.9 \pm 5$\\\hline
	\multicolumn{1}{c}{}&\multicolumn{3}{c}{$B (2)^{2}\Sigma^{+}$}\\
	\multicolumn{1}{c}{$T_e^B$} & $8933.24 \pm 0.9$&$8932.49 \pm 0.9$&$8932.53 \pm 0.9$ \\
	\multicolumn{1}{c}{$Y_{10}$} & $77.61 \pm 0.54$& $77.55 \pm0.54$& $77.56 \pm0.54$\\
	\multicolumn{1}{c}{$Y_{20}$ }& $-0.17 \pm 0.04$ & $-0.16 \pm 0.04$ & $-0.16 \pm 0.04$\\
	\multicolumn{1}{c}{$Y_{01}\times 10^{2}$ }& $3.311 \pm 0.013^{b}$& $3.397 \pm 0.014^{b}$& $3.266 \pm 0.013^{b}$\\
	\multicolumn{1}{c}{$Y_{11}\times 10^{5}$}  & $-4.8 \pm 15$& $-5.4 \pm 17$&$-8.2 \pm 25$ \\\hline	
\end{tabular} \\
\\
$^{(a)}$ Values taken from the theory and fixed during the fit.\\
$^{(b)}$ Strongly correlated with $Y_{01}$ value of the X(1)$^{2}\Sigma^{+}$ state.\\
\end{table}

It should be noted that the uncertainties of the $Y_{ij}$ values result mainly from the problem of determination of accurate positions and widths of the band heads, as many lines corresponding to transitions to different rovibrational levels in the B(2)$^{2}\Sigma^{+}$ and X(1)$^{2}\Sigma^{+}$ states overlap in the spectra and observation of only top parts of the band heads is possible. A Monte-Carlo method was used to find the error associated with this problem. With the half-width of each band head assumed somewhat arbitrarily to be 0.3 cm$^{-1}$, the positions of the individual band heads were randomly changed within the range of $\pm$0.3 cm$^{-1}$ (see Fig.~\ref{fig:bandheadformation}) and also their widths were changed within $\pm$ 0.16 cm$^{-1}$ (that influenced the number of rovibrational lines taken into consideration). A set of Dunham coefficients was determined for each random combination of the positions and widths of the band heads. The procedure was repeated until the average values of all coefficients became equal to the values determined in a straight fit described in the previous paragraph. The final errors (listed in Table \ref{tab:KSrDunham}) were defined as tripled values of the standard deviations obtained for each Dunham coefficient. Note that the $Y_{11}$ coefficients for both states are statistically meaningless but their inclusion has improved a quality of the fit.

\begin{figure}
	\includegraphics[width=0.9\linewidth]{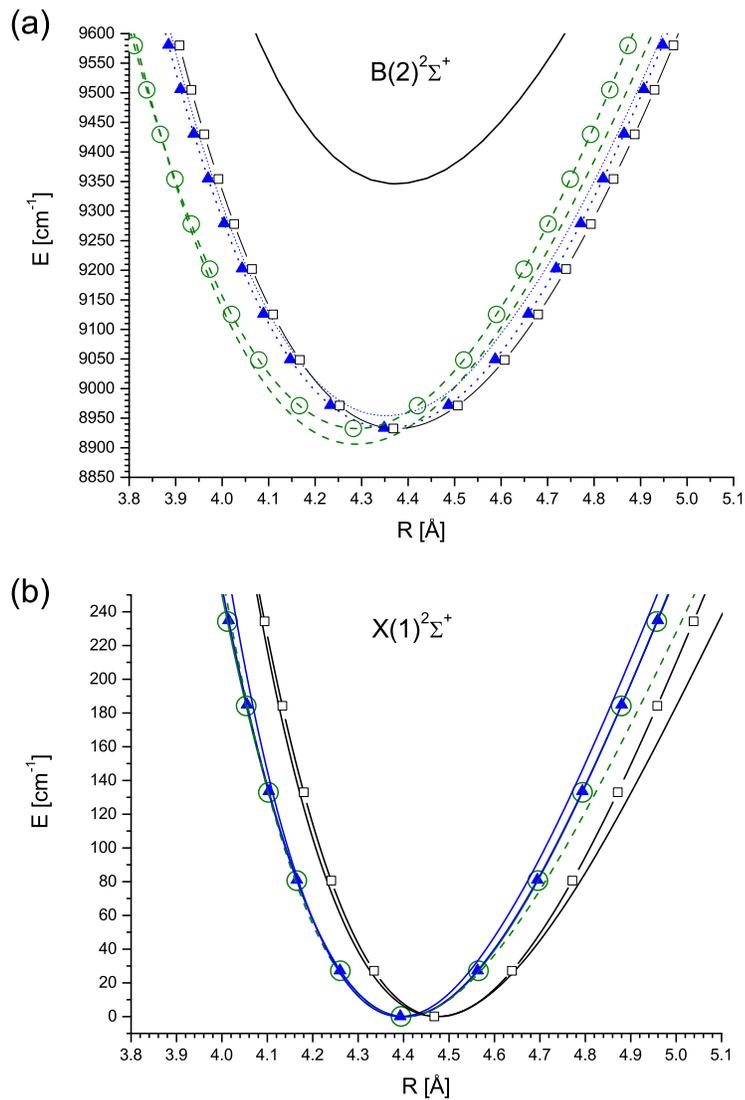}
	\caption{Comparison of the bottom parts of the theoretical curves according to the calculations with the methods FCI+ECP+CPP (dashed lines, green in colour) and RCCSD(T) (solid lines, black in colour), presented in this paper, and from the Graz group~\cite{Pototschnig} (dotted lines, blue in colour) with the experimental IPA curves of the B(2)$^{2}\Sigma^{+}$ (upper panel) and X(1)$^{2}\Sigma^{+}$ (lower panel) electronic states. The three IPA potentials for each state employ the ground state $R_e$ values predicted by the present calculations (FCI+ECP+CPP -- open circles and RCCSD(T) -- rectangles) and that reported in Ref.~\cite{Pototschnig} (triangles).}
	\label{fig:KSrIPA}
\end{figure}

Finally, three sets of rovibrational levels were generated for each of the B(2)$^{2}\Sigma^{+}$ and X(1)$^{2}\Sigma^{+}$ states basing on the experimental Dunham coefficients of Table~\ref{tab:KSrDunham}, up to 350 levels per each set, in the range of quantum numbers $v''=\left\langle 0;4\right\rangle$ , $N''=\left\langle 0;32\right\rangle$ and $v'=\left\langle 0;7\right\rangle$, $N'=\left\langle 0;33\right\rangle$. With these input data potential energy curves were built for each state using the point-wise Inverted Perturbation Approach (IPA) method \cite{Pashov2000622}. As a starting potentials the corresponding theoretical potential energy curves were used. The resulting IPA potentials are determined reliably by the experimental data in the energy range between 0 and about 250 cm$^{-1}$ for the X state and approximately between 8900 and 9550 cm$^{-1}$ for the B state. In Fig.\ref{fig:KSrIPA} the IPA potentials are compared with the bottom parts of the theoretical potential energy curves. For the theoretical potentials we assumed that the minima of the ground state curves correspond to zero energy (as in case of the experimental ones) but the calculated values of $T_e^B$ have been retained (see Table~\ref{tab:KSrconstants}). The shapes of experimental IPA curves confirm the conclusions drawn from the obtained values of the Dunham coefficients. The shape of all three experimental potentials for both states is almost identical and does not depend on the starting theoretical potential. The main difference lies in the positions of their minima and relates to the theoretical $R_{e}$ value. For the ground state positions of the potential minima differ by less than 0.1~\AA. In the case of the B(2)$^{2}\Sigma^{+}$ state the differences are even smaller. All IPA potentials are presented in the supplementary materials \cite{supplC}.

\begin{figure}
	\includegraphics[width=0.9\linewidth]{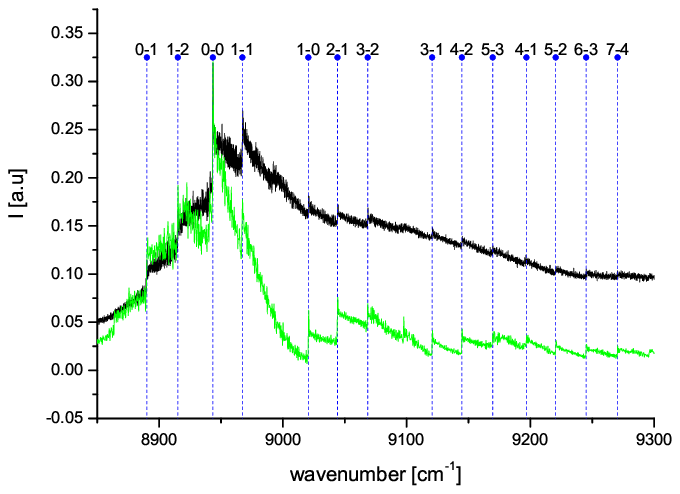}
	\caption{Comparison of the original experimental spectrum (upper curve, black in colour) with the spectrum simulated using the experimental IPA potential employing ground state $R_e$ value from the present FCI+ECP+CPP calculations (lower curve, green in colour).}
	\label{fig:IPAtest}
\end{figure}

To confirm the validity of the procedure described above, the molecular spectra were calculated again, basing on the final experimental potential curves. The exemplary result, obtained for the curves assuming the ground state $R_e$ value taken from the FCI+ECP+CPP calculations, is shown in Fig.~\ref{fig:IPAtest}. For all identified band heads there is a good agreement between their experimental and calculated positions. Moreover, the calculated spectrum allows to identify an additional band head in the observed spectrum, indicated by a question mark in Fig.~\ref{fig:expsim}(d). This proves that a good estimation of the potential energy curves of the investigated states has been obtained. 

\section{Conclusions}

With the guidance of three sets of theoretical potential energy curves and transition dipole moments - two sets being presented in this work, and one from a previous work Ref. \cite{Pototschnig} - we derived for the first time the fundamental spectroscopic constants of the KSr ground state, and of its second excited electronic state. We performed an experiment involving thermoluminescence spectroscopy, which provided molecular spectra in the X-B system with resolution 0.16~cm$^{-1}$ given by the Fourier Transform Spectrometer.

The three sets of theoretical potentials curves were used first to assign the intricate experimental spectra. Then they were employed as starting curves of the IPA routine. In each case the curves have been modified by the routine to fit the experimental data. As it turned out, the resulting potentials are of almost identical shape, irrespective of the starting curve. Only relative positions of the equilibrium internuclear distances $R_e$ differ noticeably and depend on the theoretical model, being the main limitation of our study. Nevertheless the procedure provides an interesting alternative to complicated and time-consuming experimental techniques if quick evaluation and coarse optimization of theoretical potential curves is required. Among the three theoretical approaches examined in the present work, the FCI+ECP+CPPP one seems to provide the best possible agreement with the present measurements. Further studies should be performed to confirm to which extent those three approaches could indeed be discriminated with respect to their accuracy.  We note however that their overall agreement with the derived experimental data is satisfactory, confirming that they could serve well for modeling possible schemes for ultracold molecule formation starting from ultracold atom pairs.

\section*{Acknowledgements}
J.Sz acknowledges partial support from the Miniatura I  programme founded by the National Science Centre of Poland [grant number 2017/01/X/ST2/00057].
P.S.Z  is  grateful for the Grant  OPUS 13  founded by  the National Science Centre of Poland [grant number UMO-2017/25/B/ST4/01486].
Authors acknowledge partial support from the "Hubert-Curien" partnership established between the French Ministry of High Education, Research and Innovation, of the French Ministry of Europe and Foreign Affairs, and of the Polish Ministry of Science and High Education.

\section*{References}

\bibliography{KSr_BX}

\begin{thebibliography}{10}
\expandafter\ifx\csname url\endcsname\relax
  \def\url#1{\texttt{#1}}\fi
\expandafter\ifx\csname urlprefix\endcsname\relax\def\urlprefix{URL }\fi
\expandafter\ifx\csname href\endcsname\relax
  \def\href#1#2{#2} \def\path#1{#1}\fi

\bibitem{Perez-Rios2010}
J.~{P\'{e}rez-R\'{\i}os}, F.~Herrera, R.~V. Krems, External field control of
  collective spin excitations in an optical lattice of {$^2\Sigma$} molecules,
  New J. Phys. 12 (2010) 103007.

\bibitem{Chen2014}
T.~Chen, S.~Zhu, X.~Li, J.~Qian, Y.~Wang, Prospects for transferring
  {${}^{87}$Rb${}^{84}$Sr} dimers to the rovibrational ground state based on
  calculated molecular structures, Phys. Rev. A 89 (2014) 063402.

\bibitem{Quemener2016}
G.~Qu\'em\'ener, J.~L. Bohn, Shielding $^{2}\mathrm{\ensuremath{\Sigma}}$
  ultracold dipolar molecular collisions with electric fields, Phys. Rev. A 93
  (2016) 012704.

\bibitem{Barbe2017}
V.~Barb\'{e}, A.~Ciamei, B.~Pasquiou, L.~Reichs\"{o}llner, F.~Schreck, P.~S.
  {\.{Z}uchowski}, J.~M. Hutson, Observation of {Feshbach} resonances between
  alkali and closed--shell atoms, arXiv:1710.03093.

\bibitem{Micheli2006}
A.~Micheli, G.~K. Brennen, P.~Zoller, A toolbox for lattice-spin models with
  polar molecules, Nat. Phys. 2 (2006) 341--347.

\bibitem{Baranov2012}
M.~A. Baranov, M.~Dalmonte, G.~Pupillo, P.~Zoller, Condensed matter theory of
  dipolar quantum gases, Chem. Rev. 112 (2012) 5012--5061.

\bibitem{Lahaye2009}
T.~Lahaye, C.~Menotti, L.~Santos, M.~Lewenstein, T.~Pfau, The physics of
  dipolar bosonic quantum gases, Rep. Progr. Phys. 72 (2009) 126401.

\bibitem{Lackner2014}
F.~Lackner, G.~Krois, T.~Buchsteiner, J.~V. Pototschnig, W.~E. Ernst,
  Helium-droplet-assisted preparation of cold {RbSr} molecules, Phys. Rev.
  Lett. 113 (2014) 153001.

\bibitem{Krois2014}
G.~Krois, F.~Lackner, J.~V. Pototschnig, T.~Buchsteiner, W.~E. Ernst,
  Characterization of {RbSr} molecules: spectral analysis on helium droplets,
  Phys. Chem. Chem. Phys. 16 (2014) 22373--22381.

\bibitem{Pototschnig2014}
J.~V. Pototschnig, G.~Krois, F.~Lackner, W.~E. Ernst, {\it Ab initio} study of
  the {RbSr} electronic structure: Potential energy curves, transition dipole
  moments, and permanent electric dipole moments, J. Chem. Phys. 141 (2014)
  234309.

\bibitem{Pototschnig2015}
J.~Pototschnig, G.~Krois, F.~Lackner, W.~Ernst, Investigation of the {RbCa}
  molecule: {Experiment} and theory, J. Mol. Spectrosc. 310 (2015) 126--134.

\bibitem{Krois2013}
G.~Krois, J.~V. Pototschnig, F.~Lackner, W.~E. Ernst, Spectroscopy of cold
  {LiCa} molecules formed on helium nanodroplets, J. Phys. Chem. A 117 (2013)
  13719--13731.

\bibitem{Schwanke2017}
E.~Schwanke, H.~Knoeckel, A.~Stein, A.~Pashov, S.~Ospelkaus, E.~Tiemann, Laser
  and {Fourier} transform spectroscopy of {${}^{7}$Li${}^{88}$Sr}, J. Phys. B:
  At. Mol. Opt. Phys. 50 (2017) 235103.

\bibitem{Gerschmann2017}
J.~Gerschmann, E.~Schwanke, A.~Pashov, H.~Kn\"ockel, S.~Ospelkaus, E.~Tiemann,
  Laser and {Fourier}-transform spectroscopy of {KCa}, Phys. Rev. A 96 (2017)
  032505.

\bibitem{Verges1994}
J.~Verg\`es, J.~{D'Incan}, C.~Effantin, A.~Bernard, G.~Fabre, R.~Stringat,
  A.~Boulezhar, Electronic structure of {BaLi}: the {(2)$^2\Sigma^+$} state, J.
  Phys. B: At. Mol. Opt. Phys. 27 (1994) L153--L155.

\bibitem{Stringat1994}
R.~Stringat, G.~Fabre, A.~Boulezhar, J.~{D'Incan}, V.~J. Effantin, C.,
  A.~Bernard, The {$^2\Sigma^+$}, (2){$^2\Sigma^+$} and (2){$^2\Pi$} states of
  {BaLi}, J. Mol. Spectrosc. 168 (1994) 514--521.

\bibitem{DIncan1994}
J.~{D'Incan}, C.~Effantin, A.~Bernard, G.~Fabre, R.~Stringat, A.~Boulezhar,
  J.~Verg\'{e}s, Electronic structure of {BaLi}. {II.} {First} observation of
  the {Ba$^{6,7}$Li} spectrum: Analysis of the {
  ${(2)}^{2}{\ensuremath{\Pi}}\ensuremath{\rightarrow}$
  $\mathrm{X}^{2}{\ensuremath{\Sigma}}^{+}$} system, J. Chem. Phys. 100 (1994)
  945--949.

\bibitem{russon1998}
L.~M. Russon, G.~K. Rothschopf, M.~D. Morse, A.~I. Boldyrev, J.~Simons,
  Two-photon ionization spectroscopy and all-electron {\it ab initio} study of
  {LiCa}, J. Chem. Phys. 109 (1998) 6655--6665.

\bibitem{Allouche1994}
A.~R. Allouche, M.~{Aubert-Fr\'econ}, Electronic structure of {BaLi}. {I.}
  {Theoretical} study, J. Chem. Phys. 100 (1994) 938--944.

\bibitem{boutassetta1994}
N.~Boutassetta, A.~R. Allouche, M.~Aubert-Fr\'econ, Theoretical study of the
  low-lying electronic states of the {BaNa} molecule, Chem. Phys. 189 (1994)
  33--39.

\bibitem{boutassetta1995}
N.~Boutassetta, A.~R. Allouche, M.~Aubert-Fr\'econ, Theoretical study of the
  low-lying electronic states of the {BaK} molecule, Chem. Phys. 201 (1995)
  393--403.

\bibitem{Gou2015}
D.~Gou, X.~Kuang, Y.~Gao, D.~Huo, Theoretical study on the ground state of the
  polar alkali-metal-barium molecules: Potential energy curve and permanent
  dipole moment, J. Chem. Phys. 142 (2015) 034308.

\bibitem{kotochigova2011}
S.~Kotochigova, A.~Petrov, M.~Linnik, J.~{K\l{}os}, P.~S. Julienne, {\it Ab
  initio} properties of {Li-group-II molecules} for ultracold matter studies,
  J. Chem. Phys. 135 (2011) 164108.

\bibitem{Guerout2010}
R.~Gu\'erout, M.~Aymar, O.~Dulieu, Ground state of the polar
  alkali-metal-atom\char21{}strontium molecules: Potential energy curve and
  permanent dipole moment, Phys. Rev. A 82 (2010) 042508.

\bibitem{Zuchowski2014}
P.~S. {\.{Z}uchowski}, R.~Gu\'erout, O.~Dulieu, Ground- and excited-state
  properties of the polar and paramagnetic {RbSr} molecule: A comparative
  study, Phys. Rev. A 90 (2014) 012507.

\bibitem{Gopakumar2013}
G.~Gopakumar, M.~Abe, M.~Hada, M.~Kajita, {\it Ab initio} study of ground and
  excited states of {${}^{6}$Li${}^{40}$Ca} and {${}^{6}$Li${}^{88}$Sr}
  molecules, J. Chem. Phys. 138 (2013) 194307.

\bibitem{Pototschnig2016}
J.~V. Pototschnig, A.~W. Hauser, W.~E. Ernst, Electric dipole moments and
  chemical bonding of diatomic alkali--alkaline earth molecules, Phys. Chem.
  Chem. Phys. 18 (2016) 5964--5973.

\bibitem{muller1984}
W.~M\"uller, W.~Meyer, Ground-state properties of alkali dimers and their
  cations from {\it ab initio} calculations with effective core polarization
  potentials, J. Chem. Phys. 80 (1984) 3311--3320.

\bibitem{muller1984a}
W.~M\"uller, J.~Flesch, W.~Meyer, Treatment of intershell correlation effects
  in {\it ab initio} calculations by use of core polarization potentials.
  method and application to alkali and alkaline earth atoms, J. Chem. Phys. 80
  (1984) 3297--3310.

\bibitem{foucrault1992}
M.~Foucrault, P.~Milli\'e, J.~Daudey, Nonpertubative method for {core-valence}
  correlation in pseudopotential, J. Chem. Phys. 96 (1992) 1257--1264.

\bibitem{Lim2005}
I.~Lim, P.~Schwerdtfeger, B.~Metz, H.~Stoll, All-electron and relativistic
  pseudopotential studies for the group 1 element polarizabilities from {K} to
  element 119, J. Chem. Phys. 122 (2005) 104103.

\bibitem{Lim2006}
I.~S. Lim, H.~Stoll, P.~Schwerdtfeger, Relativistic small-core
  energy-consistent pseudopotentials for the alkaline-earth elements from {Ca}
  to {Ra}, J. Chem. Phys. 124 (2006) 034107.

\bibitem{knowles1993}
P.~J. Knowles, C.~Hampel, H.~J. Werner, Coupled cluster theory for high spin,
  open shell reference wave functions, J. Chem. Phys. 99 (1993) 5219--5227.

\bibitem{Kolos2008}
J.~{K\l{}os}, M.~H. Alexander, M.~Brouard, C.~J. Eyles, F.~J. Aoiz, A new
  potential energy surface for {OH(A$^{2}\Sigma^{+}$)-Ar}: {The van der Waals}
  complex and scattering dynamics, J. Chem. Phys. 129~(5) (2008) 054301.

\bibitem{Lee2003}
T.~J. Lee, Comparison of the {T1} and {D1} diagnostics for electronic structure
  theory: a new definition for the {open-shell} {D1} diagnostic, Chem. Phys.
  Lett. 372~(3) (2003) 362 -- 367.

\bibitem{molpro2012}
H.-J. Werner, P.~J. Knowles, R.~Lindh, F.~R. Manby, M.~Sch\"{u}tz, et~al.,
  Molpro 2012-1, a package of {\it ab initio} programs (2012) see
  http://www.molpro.net/.

\bibitem{Janssen2009}
L.~M. Janssen, G.~C. Groenenboom, A.~van~der Avoird, P.~S. {\.Z}uchowski,
  R.~Podeszwa, {$Ab$ $initio$} potential energy surfaces for {NH
  (${}^3\Sigma^{-}$)--NH (${}^{3}\Sigma ^{-}$)} with analytical long range, J.
  Chem. Phys. 131 (2009) 224314.

\bibitem{Jiang2013}
J.~Jiang, Y.~Cheng, J.~Mitroy, Long-range interactions between alkali and
  alkaline-earth atoms, J. Phys. B: At. Mol. Opt. Phys. 46 (2013) 125004.

\bibitem{Pototschnig}
J.~V. Pototschnig, R.~Meyer, A.~W. Hauser, W.~E. Ernst, Vibronic transitions in
  the alkali-metal {(Li, Na, K, Rb)} \char21{} alkaline-earth-metal {(Ca, Sr)}
  series: {A} systematic analysis of {de-excitation} mechanisms based on the
  graphical mapping of {Frank-Condon} integrals, Phys. Rev. A 95 (2017) 022501.

\bibitem{supplC}
See supplementary materials at http://dx.doi.org//10.1016/j.jqsrt.2018.02.020
  and http://dimer.ifpan.edu.pl.

\bibitem{Bednarska1996}
V.~Bednarska, I.~Jackowska, W.~Jastrzebski, P.~Kowalczyk, A three-section
  heat-pipe oven for heteronuclear alkali molecules, Meas. Sci. Technol. 7
  (1996) 1291--1293.

\bibitem{Ochkin2009}
V.~Ochkin, {Spectroscopy of Low Temperature Plasma}, Wiley-VCH Verlag, 2009.

\bibitem{Pashov2000622}
A.~Pashov, W.~Jastrz\k{e}bski, P.~Kowalczyk, Construction of potential curves
  for diatomic molecular states by the {IPA} method, Comput. Phys. Commun. 128
  (2000) 622--634.

\end{thebibliography}

\end{document}